\documentclass {article}
\usepackage{color}
\usepackage{epsf}
\usepackage{epsfig}
\usepackage{graphicx}
\usepackage{latexsym}
\usepackage{amssymb}
\begin{document}

\centerline {\bf State selective differential cross sections for
single and double} \centerline {\bf electron capture in
$He\sp{1,2+}-He$ and $p-He$ collisons } \vskip 3cm \centerline {by}
\vskip 2cm \centerline  {M. S. Sch\" offler${}^1$, J. Titze${}^1$,
L. Ph. H. Schmidt${}^1$, T. Jahnke${}^1$, O. Jagutzki${}^1$, H.
Schmidt-B\" ocking${}^1$, R. D\" orner${}^1$ and I. Man\v cev${}^2$}
\vskip 2cm

\centerline {\it ${}^1$ Institut f\" ur Kernphysik, Universit\" at
Frankfurt, 60486 Frankfurt, Germany}

\centerline {\it ${}^2$ Department of Physics, Faculty of Sciences and Mathematics,}
\centerline {\it University of Ni\v s, P.O. Box 224, 18000 Ni\v s, Serbia}

\vskip 5cm \centerline {PACS: 34.70.+e - Charge transfer.}
\vfill\eject\null

{\bf Abstract.} Using the COLTRIMStechnique, scattering angle
differential cross sections for single and double electron capture
in collisions of protons and $He\sp{1,2+}$ projectiles with helium
atoms for incident energies of $60-630\,keV/u$ are measured. We also
report new theoretical results obtained by means of four-body
one-channel distorted wave models (CDW-BFS, CDW-BIS and BDW), and
find mixed agreement with the measured data.

\vskip 5mm {\bf I. Introduction} \vskip 3mm

Electron transfer is a very fundamental process in physics and
chemistry. In nuclear physics the strong forces allows protons or
neutrons being captured in nuclear collisions, in astrophysics the
gravitational force can lead to capture of stellar objects and in
chemistry capture is often a part of a chemical reaction.
Ion-atom-collisions provide an ideal testground where the quantum
mechanical electron capture processes can be explored over a wide
parameter space of trajectories (impact parameters) and velocities.

The key variable controlling the mechanism responsible for electron
transfer is the projectile velocity. At small velocities electron
transfer happens via the formation of intermediate molecular states.
At intermediate velocities capture is governed by the overlap of the
wave functions of initial and final state in momentum space,
displaced by the projectile velocity. At even higher velocities
electron capture is more likely dominated by the Thomas-process, an
inter atomic double scattering, which accelerates the target
electron to projectile velocity and leads to a distinct structure
\cite{Thomas27prsa,fischer06pra,Vogtprl,Horsdalprl,mergelprl1,mergelprl2}
in the scattering angle dependence. Finally at the highest
velocities, radiative capture dominates \cite{stoehlker94prl}.

Using the COLTRIMS-technique
\cite{ullrich97jpb,doerner00pr,ullrich03rpp} the projectile
scattering angle dependence of the electron transfer probabilities
for different final states (ground state and/or target/projectile
excitations) we measured. Especially at higher velocities only a few
experiments exist in which the different final states were
separated. In the most experiments the data have been integrated
over all final states. But for example only 30~\% of the total cross
section for single electron capture (SC) He$^{2+}$/He at 60 keV/u
lead to the ground state. Or in the most common system,
proton-Helium, (up to 630 keV/u) about 20~\% are found in an excited
state. For a quantitative comparison with theory as well as for an
interpretation of the scattering angle differential data it is,
however, essential to separate the final state. Summation of the
scattering angle dependence over excited states of projectile and or
target can wash-out much of the structure in the individual channels
\cite{mergel95prl} and obscures the comparison with theory. In this
paper we concentrate on the capture to the electronic ground state
without excitation of the remaining target electron.

\vskip 5mm {\bf II. Experiment} \vskip 3mm

Capture processes (single and multiple) typically lead to small
projectile scattering angles $\theta$, corresponding to a few atomic
units (a.~u.) of transverse momentum exchange between target and
projectile. Since no electron is emitted to the continuum, from
momentum conservation the transverse momentum transfer to the
projectile is equal and opposite to the transverse momentum of the
recoiling ion \cite{ullrich89jpb}. In the present experiment we
measure both in coincidence, the projectile scattering and the
transverse component of the recoil ion momentum.

\begin{figure}
  \epsfig{file=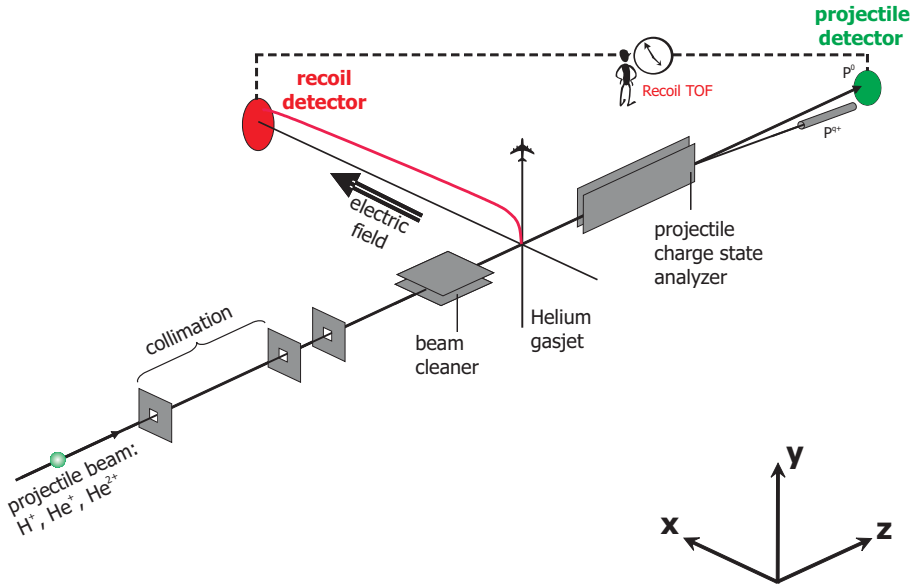,width=12.0cm}
  \caption{Draft of the experimental setup. }
  \label{figure1}
\end{figure}

The experimental data have been measured at the Van de Graaff
accelerator at the Institut f\"ur Kernphysik at the University of
Frankfurt. The final state is given by a He$^+$ recoiling ion and
the down charged ejectile. Between both collision partners momentum
and energy conservation has to be fulfilled. Thus measuring one
would be enough to get the full kinematical information. We used the
COLTRIMS technique (COLd Target Recoil Ion Momentum Spectroscopy) to
measure both particles in coincidence (see figure
\mbox{\ref{figure1}} and
\cite{ullrich97jpb,doerner00pr,ullrich03rpp} for some general
reviews).

The target is provided by a super sonic gas jet. Through a 30~$\mu$m
nozzle helium gas is expanded into a high vacuum chamber at a
driving pressure of 30~bar. 5~mm above the nozzle a skimmer
($\varnothing$ 0.3~mm) cuts out the central part of the gas jet.
Before entering the target chamber, the gas jet is collimated by a
second aperture (0.5~mm in diameter). Due to the adiabatic expansion
and the collimation the gas atoms have a momentum uncertainty of
about 0.1~a.~u. in all three dimensions. A target jet diameter of
1~mm and a density of 5$\times$10$^{11}$ atoms/cm$^2$ was reached at
the intersection region with the projectile beam. Finally the gas
jet is pumped differentially to achieve a low Helium background
pressure in the target chamber (1$\times$10$^{-8}$~mbar). The
projectile beam (H$^+$, He$^+$ and He$^{2+}$), coming from the Van
de Graaff accelerator was collimated by two sets of adjustable slits
to a beam-spot size of 0.5 $\times$ 0.5 mm$^2$ at the target. To
provide the He$^{2+}$ beam a gas stripper was used. 15~cm upstream
the target, the beam was cleaned from charge state impurities with a
set of electrostatic deflectors. Behind the target a second set of
electrostatic deflectors was used to separate the primary beam from
the ejectiles that changed their charge during the collision. The
latter were detected by a 40~mm position- and time-sensitive multi
channel plate (MCP) detector with delay line readout
\cite{jagutzki02nim}. The He$^{1+,2+}$ recoil ions were accelerated
by a weak electrostatic field of 4.8~V/cm (at the target) and
projected onto a 80~mm position- and time-sensitive MCP detector
with delay line anode for position read out. To maximize the
resolution by minimizing the perturbing influence of the extended
reaction volume a three dimensional time- and space focussing field
geometry was used (see figure 2 in \cite{doerner95nimb}). Including
drift tube, the overall distance from target to detector is 1.4~m.

By measuring the time of flight (19~$\mu$s for He$^+$ and
13.4~$\mu$s for He$^{2+}$) we obtained the charge state and the
momentum in field direction. From the position of impact we
calculated the momenta in the direction perpendicular to the
electric field.

Along the beam axis (z) the momentum of the recoil ion is directly
related to the Q-value of the reaction \cite{mergel95prl}:

\begin{eqnarray}
    p_{z} = -\frac{Q}{v_P} - \frac{v_P}{2}
\end{eqnarray}

Each final electronic state corresponds to a well defined discrete
longitudinal ion momentum. In reality these peaks are broadened by
the target temperature and the resolution of the spectrometer. Their
width is the total momentum resolution, which was found to be
0.1~a.~u. (limited by the target temperature). The measured
experimental values for all final states have been normalized to
total cross sections, taken from \cite{Barnett90}. The projectile
scattering angles was also measured in coincidence. It was used,
however, only to roughly clean the data from background (by checking
for momentum conservation in the plane perpendicular to the initial
beam axis). Instead the projectile scattering angle $\theta$ was
deduced from the momenta transferred to the target, for which we
achieve a much better resolution than for the projectile momentum
itself. The spectrometer's geometry and voltages were chosen to
yield 4$\pi$ acceptance angle for ions up to 10~a.~u. transverse
momentum.

\vskip 5mm {\bf III. Theory} \vskip 3mm

{\bf A. Single electron capture}

  Let us first consider single charge exchange in collisions of completely
  stripped projectiles with a helium-like target: $Z_P+(Z_T;e_1,e_2)_i
  \longrightarrow (Z_P;e_1)_{f_1}+(Z_T;e_2)_{f_2},$ where $Z_P(Z_T)$ is
  the charge of the projectile (target).  The parentheses symbolize the bound
  state whose quantum numbers are given by collective labels $i$, $f_1$ and
  $f_2$. We shall denote by $\vec s_{1,2}$ and $\vec x_{1,2}$ the position
  vectors of the electrons $e_{1,2}$ relative to $Z_P$ and $Z_T$,
  respectively. The inter-electron distance $r_{12}$ is given by
  $r_{12}=|\vec s_1-\vec s_2|=|\vec x_1-\vec x_2|$. Let further $\vec R$
  be the vector of the internuclear axis directed from $Z_T$ to $Z_P$.
  In the entrance channel, it is convenient to introduce $\vec r_i$
  as a relative vector of $Z_P$ with respect to the center of mass of
  $(Z_T;e_1,e_2)_i.$ Further in the exit channel, let $\vec r_f$ be the
  relative vector of the center of mass of $(Z_P;e_1)_{f_1}$ with
  respect to the center of mass of $(Z_T;e_2)_{f_2}$.

 The asymptotic channel states $\Phi_{i,f}^\pm$ which satisfy correct
boundary conditions read as follows:
\begin{eqnarray}
\Phi_i^+=\varphi_i(\vec x_1,\vec x_2)\exp[i\vec k_i\cdot \vec
r_i+i\nu_i\ln (k_ir_i-\vec k_i\cdot \vec r_i)],\label {eq1}
\end{eqnarray}
\begin{eqnarray}
 \Phi_f^-=
\varphi_P(\vec s_1)\varphi_T(\vec x_2)
\exp[-i\vec k_f\cdot\vec r_f-i\nu_f\ln(k_fr_f- \vec k_f\cdot\vec
r_f)],\label{eq2}
\end{eqnarray}
where $\nu_i=Z_P(Z_T-2)/v$, $\nu_f=(Z_T-1)(Z_P-1)/v$ with $v$ being
the incident velocity, and $k_i, k_f$  are  the  initial and final
wave vectors. The initial bound state is denoted by $\varphi_i(\vec
x_1,\vec x_2)$, whereas bound state wave functions of the atomic
system $(Z_P;e_1)$ and $(Z_T;e_2)$ are labeled by $\varphi_P(\vec
s_1)$ and $\varphi_T(\vec x_2)$, respectively.

The transition amplitudes in the continuum distorted wave - Born
initial state (CDW-BIS) approximation \cite{man4,man3}
 and the continuum distorted wave - Born
final state (CDW-BFS) \cite{man1,man2} are given as:
\begin{eqnarray}
T^{CDW-BIS}_{if}=\langle \chi_f^-|U_f^\dagger|\Phi_i^+\rangle,
\qquad
T^{CDW-BFS}_{if}=\langle \Phi_f^-|U_i|\chi_i^+\rangle.
\label{eq3}
\end{eqnarray}

Here, the  perturbation potentials $U_i$ and $U_f$  and corresponding
distorted waves $\chi_i^\pm$ are chosen as in the CDW-4B method
\cite{cdw1,cdw2}:
\begin{eqnarray}
U_i=V(R,s_2)-\vec \nabla_{x_1}\ln \varphi_i(\vec x_1,\vec x_2)\cdot
\vec \nabla_{s_1},\label{eq4}
\end{eqnarray}
\begin{eqnarray}
U_f=V(R,s_2)-V(r_{12},x_1)-\vec \nabla_{s_1}\ln \varphi_f(\vec s_1)\cdot
\vec \nabla_{x_1},\label{eq5}
\end{eqnarray}
with
\begin{eqnarray}
V(R,s_2)=Z_P\left(1/R- 1/{s_2}\right),\quad
V(r_{12},x_1)=\left( 1/{x_1}- 1/r_{12}\right),\label{eq6}
\end{eqnarray}
\begin{eqnarray}
&& \chi_i^+=N^+(\nu_P){\cal N}^+(\nu)e^{i \vec k_i\cdot\vec r_i}
 \varphi_i(\vec x_1,\vec x_2){}_1F_1(i\nu_P,1,ivs_1+i\vec v\cdot \vec
s_1)\nonumber\\
&&\hskip 5mm\times  {}_1F_1(-i\nu,1,ik_ir_i-i\vec k_i\cdot \vec r_i),
\label{eq7}
\end{eqnarray}
\begin{eqnarray}
&&\chi_f^-= {\cal N}^-(\nu)N^-(\nu_T)  \varphi_P(\vec s_1)\varphi_T(\vec
x_2) e^{-i\vec k_f\cdot \vec r_f}   {}_1F_1(-i\nu_T,1,-ivx_1-i\vec
v\cdot\vec x_1)\nonumber \\
 &&\hskip 5mm \times {}_1F_1(i\nu,1,-ik_fr_f+i\vec k_f\cdot\vec r_f),
\label{eq8}
\end{eqnarray}
where
$\nu_T=(Z_T-1)/v,$ $\nu_P=Z_P/v,$ $\nu=Z_P(Z_T-1)/v,$
 $N^-(\nu_T)=\Gamma(1+i\nu_T)e^{\pi\nu_T/2},$
 $N^+(\nu_P)=\Gamma(1-i\nu_P) e^{\pi \nu_P/2},$
 ${\cal N}^\pm(\nu)=\Gamma(1\pm i\nu)e^{-\pi\nu/2}.$
 The symbol ${}_1F_1(a,b,c)$ denotes the confluent hypergeometric
function.  The eikonal approximation $\vec R\simeq -\vec r_f$, $\vec R\simeq
\vec r_i$ is also used.
 It is  readily verified that the distorted  waves
$\chi_{i,f}^\pm$ satisfy the proper boundary conditions
$\chi_{i,f}^\pm\to \Phi_{i,f}^\pm$ when $r_{i,f}\to \infty$.

 The  explicit  expressions for matrix elements  can be  written as:
\begin{eqnarray*}
&&\kern -10mm T^{CDW-BFS}_{if}\!=\!N^+(\nu_P)\!\int\!\int\!\int d\vec
x_1d\vec x_2d\vec R e^{i \vec k_i\cdot\vec r_i
+i \vec k_f\cdot\vec r_f}\varphi^*_P(\vec s_1)\varphi^*_T(\vec x_2){\cal
L}(R)  \left[  V(R,s_2) \right.\nonumber\\
&&\kern -10mm \times  \left.\varphi_i(\vec x_1,\vec x_2)
{}_1F_1(i\nu_P,1,ivs_1\!+\!i\vec v\!\cdot\! \vec s_1)
\!-\!\vec\nabla_{x_1}\varphi_i(\vec x_1,\vec x_2)\!\cdot\!
\vec\nabla_{s_1}{}_1F_1(i\nu_P,1,ivs_1\!+\!i\vec v\!\cdot\! \vec s_1)\right],
\end{eqnarray*}

\begin{eqnarray*}
&& T^{CDW-BIS}_{if} =[N^-(\nu_T)]^*\int\int\int d\vec x_1d\vec x_2d\vec
R e^{i \vec k_i\cdot\vec r_i +i \vec k_f\cdot\vec r_f}\varphi_i(\vec
x_1,\vec x_2) {\cal R} (R)\nonumber\\
&&   \left\{{}_1F_1(i\nu_T,1,ivx_1+i\vec v\cdot \vec x_1)\right.
  [V(R,s_2)\!-\!V(r_{12},x_1)]\varphi^*_P(\vec
s_1)\varphi^*_T(\vec x_2)\nonumber\\
&&\left. - \varphi^*_T(\vec x_2)\vec\nabla_{s_1}\varphi^*_P(\vec s_1)\!\cdot \!
\vec\nabla_{x_1}{}_1F_1(i\nu_T,\!1,ivx_1+ i\vec v\cdot \vec
x_1)\!\right\},
\end{eqnarray*}
with
\begin{eqnarray}
{\cal L}(R)\!=\! (vR\!+\!\vec v\!\cdot\! \vec R)^{i\nu_f}
(vR\!-\!\vec v\!\cdot\! \vec R)^{i\nu}
\!=\! \rho^{2iZ_P(Z_T-1)/v}(vR\!+\!\vec v\!\cdot\! \vec R\,)^{-i(Z_T-1)/v },
\label{eq11}
\end{eqnarray}
\begin{eqnarray}
{\cal R}(R)= (vR-\vec v\cdot \vec R)^{i\nu_i}
(vR+\vec v\cdot \vec R)^{i\nu} =(\rho )^{2i\nu_i} (vR+\vec v\cdot
\vec R)^{iZ_P/v},\label{eq12}
\end{eqnarray}
where unimportant phase factors are dropped.
 Here $\vec \rho$ is
a component of the vector of the internuclear distance
perpendicular to the $Z$-axis.

The analytical calculation outlined in the ref. \cite{man1} provides
the matrix elements of the CDW-BFS model in terms of two dimensional
real quadrature, whereas in the case of CDW-BIS approximation the
$T$-matrix elements can be analytically reduced \cite{man3,man4} to
five-dimensional integral which must be evaluated numerically.
Reason for this is that the term $1/r_{12}$ in the perturbation
$U_f$ requires an additional three-dimensional integral.  All
numerical integrations are carried out by means of the
Gauss-Legendre  quadrature after scaling of variables. In both
models, the standard Cauchy regularization of the whole integrand is
accomplished before applying the numerical integrations.

 During the construction of the hybrid-type  models, such as CDW-BIS and
 CDW-BFS, the main idea has been to approximate the exact wave function
{\it in one} of the channels, by using a simple analytical function
which can well describe the principal interaction region, and, to
preserve correct boundary conditions in both channels. Hence,
according to these models  the captured electron is treated in an
asymmetrical manner in the entrance and exit channel.

In this work, the explicit calculation of the matrix elements for
single electron capture are carried out by using the
 two-parameter wave function of Silverman {\it et al} \cite{Silverman}
for the initial state of helium  target:
 $\varphi_i(\vec x_1,\vec x_2)=N[e^{-\alpha_1 x_1-\alpha_2 x_2}+e^{-\alpha_2 x_1-\alpha_1 x_2}]/\pi   $,
where  $ N=\left[1/{\alpha_1^3}+1/{\alpha_2^3}+
16/(\alpha_1+\alpha_2)^3\right]^{-1/2}.$
 Despite  its very simple form in this function \cite{Silverman} the radial
static correlations are  taken into account to within nearly 90\%.

{\bf B. Double electron capture}

Next, we consider typical double charge exchange in collisions of
completely stripped projectiles with heliumlike target:
 $Z_P+(Z_T;e_1,e_2)_i  \longrightarrow (Z_P;e_1,e_2)_f+ Z_T.$ In the
 present paper the second order four-body theory called Born distorted
wave (BDW) approximation \cite{DC1,DC2} is employed.
 According to this model the transition amplitude (for example prior
form) is given by:
 \begin{eqnarray}
 T^{BDW}_{if}=\langle \Phi_f^-|U_i|\chi_i^+\rangle,
\label{eq13}
\end{eqnarray}
where the asymptotic channel state $\Phi_f^-$ reads as follows:
\begin{eqnarray}
 \Phi_f^-=
\varphi_f(\vec s_1,\vec s_2)  \exp[-i\vec k_f\cdot\vec
r_f-i\nu_f\ln(k_fr_f- \vec k_f\cdot\vec r_f)],
\label{eq14}
\end{eqnarray}
with $\nu_f=Z_T(Z_P-2)/v$. Now the $\vec r_f$   denotes relative
vector of $Z_P$ with respect to the center of mass of
$(Z_T;e_1,e_2)_f$ in the exit channel. The initial distorted wave
$\chi_i^+$  is given in the eikonal limit:

\begin{eqnarray}
&& \chi_i^+=[N^+(\nu_P)]^2{\cal N}^+(\nu_{PT})e^{i \vec k_i\cdot\vec r_i}
 \varphi_i(\vec x_1,\vec x_2){}_1F_1(i\nu_P,1,ivs_1+i\vec v\cdot \vec
s_1)\nonumber\\
&& {}_1F_1(i\nu_P,1,ivs_2+i\vec v\cdot \vec
s_2)  {}_1F_1(-i\nu_{PT},1,ik_ir_i-i\vec k_i\cdot \vec r_i),
\label{eq15}
\end{eqnarray}
provided that $U_i$ is chosen as
\begin{eqnarray}
U_i=\sum_{i=1}^{2}\vec \nabla_{x_i}\ln \varphi_i(\vec x_1,\vec
x_2)\cdot \vec \nabla_{s_i}. \label{eq16}
\end{eqnarray}
Here, the symbols $\nu_{PT}$ and $\nu_P$ are defined as:
$\nu_{PT}=Z_PZ_T/v,$ $\nu_P=Z_P/v$.
 Hence, the BDW  is hybrid-type  model which in the entrance and
exit channel coincides, respectively, with the CDW-4B \cite{DC3} and
CB1-4B \cite{DC4} method.
 The BDW approximation introduces a normalized scattering state for
every value of $r_f$ in the final channel, in which the reduction of
the BDW to the CB1-4B model takes place.

The differential cross sections are defined by the following relation
\begin{eqnarray}
\frac {d\sigma d\Omega}=\frac
{\mu_i\mu_f}{4\pi^2}\left|T_{if}\right|^2 \,\,[a_0^2/sr]
\label{eq17}
\end{eqnarray}
where $\mu_i=M_P(M_T+2)/(M_P+M_T+2)$, $\mu_f=M_T(M_P+2)/(M_P+M_T+2)$
with $M_{P,T}$ being mass of the projectile/target nucleus.
 The scattering angle $\theta$ is identified from the relation
$\eta=2\mu_{PT}v\sin(\theta/2)$, where $\mu_{PT}$ is the reduced mass
of the projectile and target nucleus.

\vskip 5mm {\bf IV. Results and discussion} \vskip 3mm

\centerline{\it p+He collisions} \vskip 2mm

\begin{figure}
  \epsfig{file=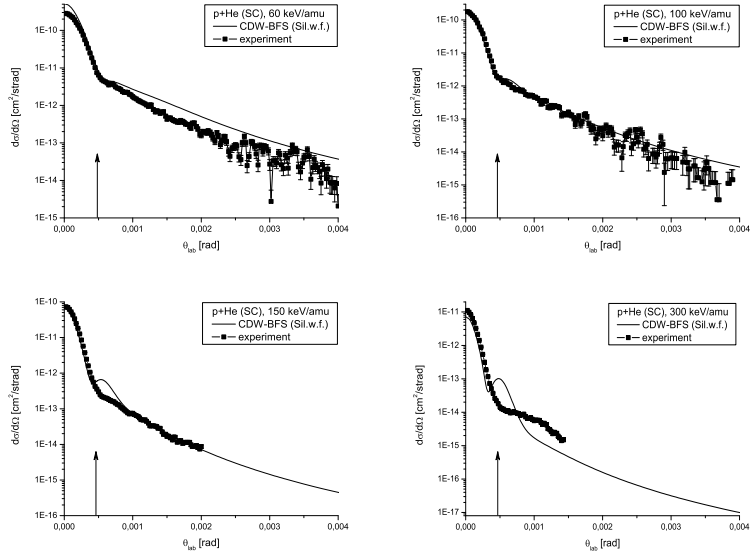,width=12.0cm}
  \caption{The differential cross sections $dQ_{if}/d\Omega(cm^2/strad)$ as a
           function of scattering angle $\theta(rad)$ for single electron capture
           to the groundstate in $p-He$ collisions at $60\,$, $100\,$, $150\,$ and $300\,keV/u$.
           Both cross sections and the scattering angle are in the laboratory system.
           The symbol $\kern -0.5mm {\vrule width 1.7mm height 1.7mm \,}$ relates
           to the present measurements, the full line represents the results
           obtained by means of the CDW-BFS model (present computation). The
           arrow indicated the location of the electron-nuclear Thomas-peak. }
  \label{figure2}
\end{figure}

Theoretical results for differential cross sections for single
electron capture in $p-He$ collisions at impact energies $60$,
$100$, $150$, $300$ keV obtained by means of four body CDW-BFS model
are compared with the measured data in Figure
\mbox{\ref{figure2}}a-d. The displayed theoretical results in Fig.
\mbox{\ref{figure2}} are obtained by means of a two parameter
Silverman {\it et al.} \cite {Silverman} orbitals for the initial
helium bound state. The computations are performed also with the
help of simplest one parameter orbitals of Hylleraas \cite{Hyll} as
well as four-parameter wave functions of L\" owdin \cite{Lowdin}. It
is observed that the differential cross section for the here
considered problem is not strongly sensitive to the choice of the
bound state wave functions, since the difference between
corresponding results is less than $20\%$. In general the main peak
in the differential cross sections between 0 and 0.5 mrad is
mediated by the momentum transfer of the captured electron and
reflects an image of the initial electron momentum distribution.
Scattering angles below 0.55~mrad are known to be dominated by
momentum transfer mediated by the electron (see
\cite{Kamber88prl,doerner89prl,gensmantel92pra} and
\cite{mergelprl1} for capture reactions) while all larger deflection
angles the scattering is dominated by momentum exchange between the
nuclei.

The theory yields good agreement as well as in shape as in absolute
height with the experimental data at small scattering angles for all
energies, also for other measurements \cite{Bross,Martin}. However,
for larger scattering angles theory predcits a slightly different
angular and $v_P$-dependency. Particularly the CDW-BFS-approximation
predicts a strong Thomas-peak contribution
($\theta_{lab}=(1/M_P)\sin 60^o\simeq 0.472\, mrad$) with increasing
$v_P$, which is clearly not present in the data. At impact energies
of 300 keV/u theory increasingly underestimates the large scattering
angles clearly..

\vskip 3mm \centerline{\it ${}^3He^{2+}+{}^4He$ collisions (Single
capture)} \vskip 2mm

\begin{figure}
  \epsfig{file=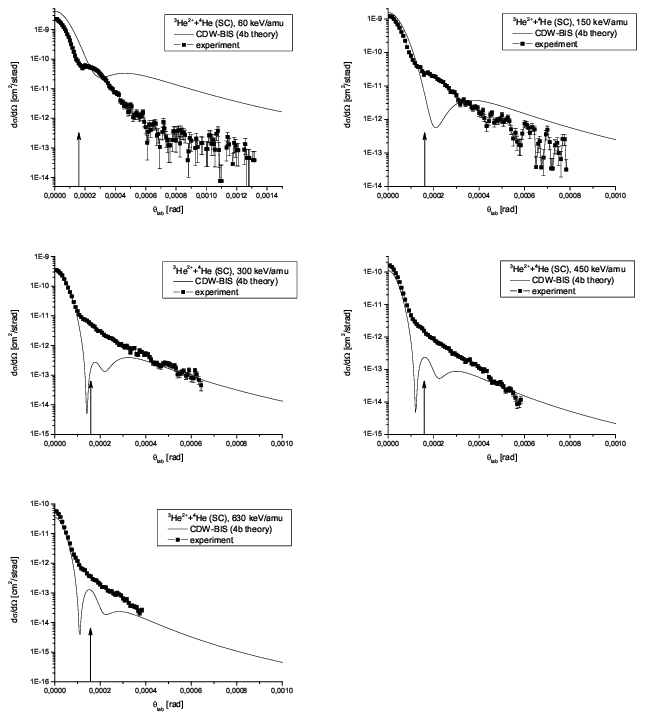,width=12.0cm}
  \caption{The differential cross sections $dQ_{if}/d\Omega(cm^2/strad)$ as a
           function of scattering angle $\theta(rad)$ for single electron capture
           to the groundstate in $He^{2+}-He$ collisions at $60\,$, $150\,$, $300\,$ $450\,$ and $630\,keV/u$.
           Both cross sections and the scattering angle are in the laboratory  system.
           The symbol $\kern -0.5mm {\vrule width 1.7mm height 1.7mm \,}$ relates
           to the present measurements, the solid line represents the results
           obtained by means of the CDW-BIS model (present computation). The
           arrow indicated the location of the electron-nuclear Thomas-peak. }
  \label{figure3}
\end{figure}

Theoretical results for single electron capture in ${}^3He^{2+}+He$
collisions at impact energies $60$, $150$, $300$, $450$ and
$630\,keV/u$ for angular region from $0$ to $1.5$~mrad obtained by
means of four body CDW-BIS model are compared with the measured data
in Figures \mbox{\ref{figure3}}. The Thomas peak for this reaction
is at $\theta_{lab} \simeq 0.154\, mrad$. The CDW-BIS model exhibits
an unphysical and experimentally unobserved dip before and after the
Thomas peak region, due to mutual cancelation among the various
terms in the potential $U_f$ given by eq.(\ref{eq5}). Despite the
proper inclusion of the Rutherford scattering, the CDW-BIS
approximation predicts the differential cross sections which are in
disagreement with measurements in the region where the internuclear
scattering takes over. Nevertheless, the experimental resolution is
clearly good enough to make such structures visible, if they would
exist. So thus at $60\,keV/u$ the theoretical curve is considerably
higher than the corresponding experimental data at larger scattering
angle.

\vskip 3mm \centerline{\it ${}^3He^{+}+{}^4He$ collisions (Single
capture)} \vskip 2mm

\begin{figure}
  \epsfig{file=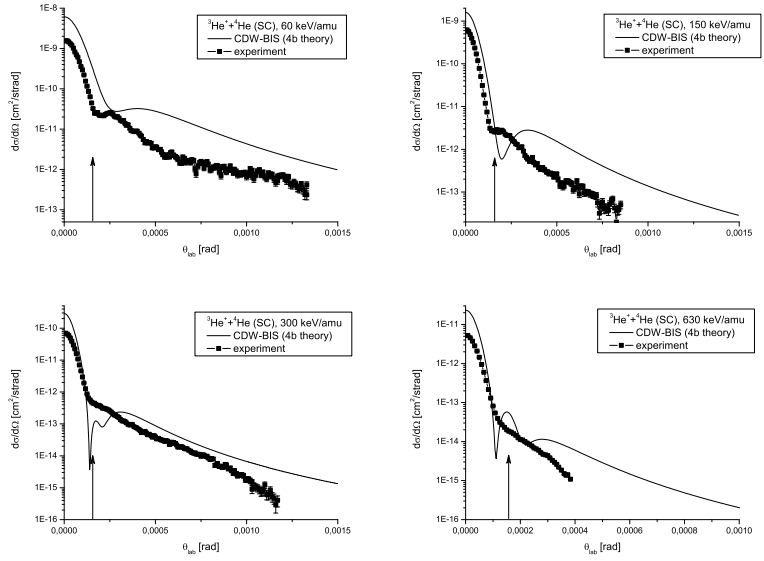,width=12.0cm}
  \caption{The differential cross sections $dQ_{if}/d\Omega(cm^2/strad)$ as a
           function of scattering angle $\theta(rad)$ for single electron capture
           to the groundstate in $He^{+}-He$ collisions at $60\,$, $150\,$, $300\,$ and $630\,keV/u$.
           Both cross sections and the scattering angle are in the laboratory  system.
           The symbol $\kern -0.5mm {\vrule width 1.7mm height 1.7mm \,}$ relates
           to the present measurements, the solid line represents the results
           obtained by means of the CDW-BIS model (present computation). The
           arrow indicated the location of the electron-nuclear Thomas-peak. }
  \label{figure4}
\end{figure}

This is a true five-body problem of the type
$(Z_P;e_3)+(Z_T;e_1,e_2)_i  \longrightarrow (Z_P;e_1,e_3)_{f_1}+
(Z_T;e_2)_{f_2}.$ Instead of this, in the present work we
theoretically consider the following model reaction:
$Z_P^{eff}+(Z_T;e_1,e_2)_i  \longrightarrow (Z_P^{eff};e_1)_{f_1}+
(Z_T;e_2)_{f_2}.$ In this model the presence of the projectile
electron is taken into account only through a screening effect. The
two-electron atom in the exit channel is described by a hydrogenic
model. Within this approximation, the binding energy is
$E_f=-[Z_P^{eff}]^2/2$, where the effective charge is
$Z_P^{eff}=Z_P-5/16$. Such a choice of the $Z_P^{eff}$ provides a
satisfactory description for a $K$-shell electron near the Bohr
radius. The corresponding hydrogenic $1s$ wave function is very
close to the actual $K$-shell orbital (for example Hartree-Fock $1s$
wave function). However, this fixed effective charge does not
produce the correct experimental binding energy. It also cannot
reflect the true dynamic situation of the captured electron in the
newly formed helium atom. Of course, other choices for the effective
charge are possible (see e.g. \cite{McGuire}). The results of the
computation, using CDW-BIS theory, of such a model reaction are
represented in Fig. \mbox{\ref{figure4}}a-d.

It can be seen from figure~\mbox{\ref{figure4}}, that this
satisfactory theoretical approach cannot reproduce the data. At
larger scattering angle we find a similar disagreement as for the
collision systems discussed previously. But now also in the small
angle regime ("the electronic peak") the $v_P$ dependence is not
well predicted. Similar to the impact of $He^{2+}$ the theory shows
an unphysical dip around the Thomas peak.

\vskip 3mm \centerline{\it ${}^3He^{2+}+{}^4 He$ collisions (Double
capture)} \vskip 2mm

\begin{figure}
  \epsfig{file=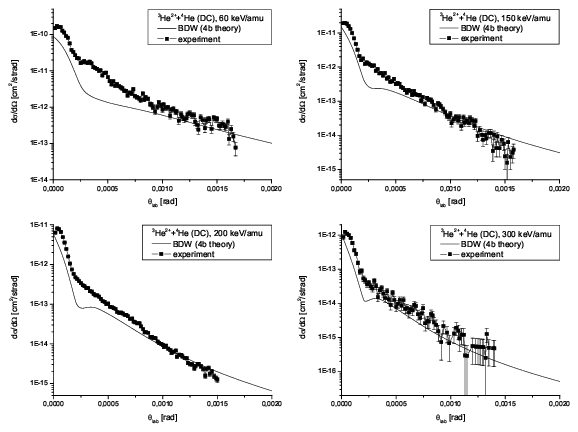,width=12.0cm}
  \caption{The differential cross sections $dQ_{if}/d\Omega(cm^2/strad)$ as a
           function of scattering angle $\theta(rad)$ for double electron capture
           to the groundstate in $He^{+}-He$ collisions at $60\,$, $150\,$, $200\,$ and $300\,keV/u$.
           Both cross sections and the scattering angle are in the laboratory  system.
           The symbol $\kern -0.5mm {\vrule width 1.7mm height 1.7mm \,}$ relates
           to the present measurements, the full line  represents the results
           obtained by means of the BDWS approximation (present computation).}
  \label{figure5}
\end{figure}

Theoretical results for differential cross sections in
${}^3He^{2+}+He$ collisions  at incident energies $60$, $150$, $200$
and $300\,keV/u$ are displayed in Fig. \mbox{\ref{figure5}}a-d. As
can be seen, the BDW-calculations are in general in the
electron-peak regime a factor 2 below the experimental data, but
agree well in shape. At larger scattering angles (nuclear
scattering), experimental and theoretical slope differ slightly. The
displayed theoretical and experimental results present only the
transition $1s^2\to 1s^2$. The present computations for double
capture are performed by using one-parameter orbitals of the
Hylleraas type \cite{Hyll} in both the entrance and exit channel.
For example, the initial state bound state of the helium target is
described  by $\varphi_i(\vec x_1,\vec
x_2)=(\gamma^3/\pi)exp[-\gamma(x_1+x_2)]$, where $\gamma=1.6875$ is
the effective charge.

The computations of differential cross sections are carried out also
by means of the four-body boundary-corrected continuum intermediate
state (BCIS) approximation which has been introduced by Belki\' c
\cite{DC5}. The essential difference between the BCIS and BDW is in
the perturbation potentials. Namely,  in the BDW model one
encounters the typical gradient operator potentials which are
familiar from the CDW-4B approximation \cite{DC3}. The role of these
perturbations in the BCIS method is played by the conventional first
Born-type multiplicative operators, which are the same as in the
CB1-4B approach \cite{DC4}. It is observed that the behavior of the
angular distribution obtained in the BCIS model is altogether quite
similar to that in the BDW approximation for the considered impact
energies.

\vskip 3mm \centerline{\it Occupied states} \vskip 2mm

As explained from the experimental values we can deduce the fraction
of different occupied final states (transfer to groundstate,
target-, projectile- and target\&projectile excitation). For the
impact He$^{2+}$ it couldn't be distinguished whether the target
(He$^+$) or the projectile (either He$^+$) is excited. Hence the sum
of both is presented in table \mbox{\ref{table1}} below.

\begin{table}[htb]
  \begin{center}
    \begin{tabular}{|l|c|c|c|c|}
    \hline
    collision system & $\sigma (P_{1s}, T_{1s})$ & $\sigma (P_{nl}, T_{nl})$ & $\sigma (P_{nl}, T_{1s})$ & $\sigma (P_{nl}, T_{nl})$ \\ \hline
    $H^+$ 60 keV/u SC & $6.5\times10^{-17}$ & $2.3\times10^{-18}$ & $1.5\times10^{-17}$ & $1.2\times10^{-18}$ \\
    $H^+$ 100 keV/u SC & $2.5\times10^{-17}$ & $1.0\times10^{-18}$ & $6.8\times10^{-18}$ & $6.5\times10^{-19}$ \\
    $H^+$ 150 keV/u SC & $7.3\times10^{-18}$ & $3.0\times10^{-19}$ & $2.0\times10^{-18}$ & $1.4\times10^{-19}$ \\
    $H^+$ 300 keV/u SC & $7.0\times10^{-19}$ & $4.6\times10^{-20}$ & $1.6\times10^{-19}$ & $1.0\times10^{-20}$ \\ \hline
    $He^{2+}$ 40 keV/u SC & $7.9\times10^{-17}$ & \multicolumn{2}{|c|}{$2.2\times10^{-16}$} & $3.0\times10^{-18}$ \\
    $He^{2+}$ 60 keV/u SC & $5.7\times10^{-17}$ & \multicolumn{2}{|c|}{$1.5\times10^{-16}$} & $4.4\times10^{-18}$ \\
    $He^{2+}$ 150 keV/u SC & $1.8\times10^{-17}$ & \multicolumn{2}{|c|}{$2.2\times10^{-17}$} & $2.7\times10^{-18}$ \\
    $He^{2+}$ 300 keV/u SC & $3.9\times10^{-17}$ & \multicolumn{2}{|c|}{$3.4\times10^{-18}$} & $4.3\times10^{-19}$ \\
    $He^{2+}$ 450 keV/u SC & $1.4\times10^{-17}$ & \multicolumn{2}{|c|}{$9.0\times10^{-19}$} & $7.4\times10^{-20}$ \\
    $He^{2+}$ 630 keV/u SC & $3.4\times10^{-17}$ & \multicolumn{2}{|c|}{$1.9\times10^{-19}$} & $2.3\times10^{-20}$ \\ \hline
    $He^+$ 60 keV/u SC & $4.2\times10^{-17}$ & $1.9\times10^{-17}$ & $9.1\times10^{-18}$ & $8.5\times10^{-19}$ \\
    $He^+$ 150 keV/u SC & $5.9\times10^{-18}$ & $3.1\times10^{-18}$ & $6.4\times10^{-19}$ & $3.7\times10^{-19}$ \\
    $He^+$ 300 keV/u SC & $7.9\times10^{-19}$ & $2.9\times10^{-19}$ & $7.9\times10^{-20}$ & $3.2\times10^{-20}$ \\
    $He^+$ 630 keV/u SC & $3.5\times10^{-20}$ & $1.1\times10^{-20}$ & $3.2\times10^{-21}$ & $1.1\times10^{-21}$ \\ \hline
    $He^{2+}$ 40 keV/u DC & $4.5\times10^{-17}$ & - & $1.6\times10^{-17}$ & - \\
    $He^{2+}$ 60 keV/u DC & $2.7\times10^{-17}$ & - & $7.5\times10^{-18}$ & - \\
    $He^{2+}$ 150 keV/u DC & $1.7\times10^{-18}$ & - & $3.5\times10^{-19}$ & - \\
    $He^{2+}$ 200 keV/u DC & $5.3\times10^{-19}$ & - & $1.1\times10^{-19}$ & - \\
    $He^{2+}$ 300 keV/u DC & $7.3\times10^{-20}$ & - & $1.2\times10^{-20}$ & - \\
    \hline
    \end{tabular}
    \caption{Total cross sections for single and double electron transfer
             to the ground state $\sigma (P_{1s}, T_{1s})$, with target excitation $\sigma (P_{nl}, T_{nl})$,
             with projectile excitation $\sigma (P_{nl}, T_{1s})$ and
             target+projectile excitation $\sigma (P_{nl}, T_{nl})$,
             $n,l\geq2$. The integral data have been normalized to \cite{Barnett90}. }
    \label{table1}
 \end{center}
\end{table}

\vskip 5mm {\bf V. Conclusions} \vskip 3mm

In conclusion we have presented a systematic comparison of
experimental data and theory for fully differential data of the most
simple capture reactions in an intermediate range of projectile
energies. At very small angles (the dominant contribution to the
total capture cross section) the electron-transfer is mediated by
the electron dynamics in the initial state, yielding an universal
shape for all systems investigated. Theory and experiment agree
quite well in shape and absolute height (beside double capture). At
large scattering angles (nuclear scattering), we observe larger
disagreement between theory and present measurement. These findings
are in line with similar conclusion drawn in the last few years for
ionization collisions
\cite{Schulznature,olson2006jpb,foster2006prl}. Also for ionization
surprising discrepancies between experiment and the most elaborated
theoretical approaches persist. They are attributed currently to
problems in describing correctly the nuclear and electron momentum
exchange \cite{Schulz2007pra}. The delicate interplay of momentum
exchange between nuclei and electrons in few body collision still
poses a major challenge to theory today.

{\bf Acknowledgements}

This work was supported by DFG, the BMBF, GSI, and Roentdek GmbH.
I.M. acknowledges the support from Ministry of Science of the
Republic of Serbia through Project No. 141029A.

{}

\end{document}